\documentclass[american,aps,showpacs,showkeys]{revtex4}
\usepackage[latin9]{inputenc}
\usepackage{dcolumn}
\usepackage{color}
\usepackage{textcomp}
\usepackage{amsmath}
\usepackage{amssymb}
\usepackage{graphicx}
\usepackage{esint}

\makeatletter
\usepackage{dcolumn}
\usepackage{bm}
\usepackage{wrapfig}
\usepackage{subfigure}
\usepackage{ucs}
\usepackage{lineno}
\usepackage{epsfig}
\usepackage{psfrag}\usepackage{epstopdf}
\DeclareGraphicsExtensions{.pdf,.eps,.png,.jpg,.mps}
\usepackage{babel}
\makeatother
\usepackage{babel}
\begin{document}

\title{Magnetic behavior of a ferro-ferrimagnetic ternary alloy AB$_\rho$C$_{1-\rho}$ with a selective site disorder: the case study of a mixed-spin Ising model on a honeycomb lattice}

\author{Jordana Torrico$^{1}$, Jozef Stre\v{c}ka$^{2}$, Onofre Rojas$^{1}$, Sergio Martins de Souza$^{1}$, Marcelo Leite Lyra$^{3}$}

\affiliation{$^{1}$Departamento de F\'isica, Universidade Federal de Lavras, 37200-000, Lavras-MG}

\affiliation{$^{2}$Department of Theoretical Physics and Astrophysics, Faculty of Science,
P. J. $\check{S}$af\'arik University, Park Angelinum 9, 040 01 Ko$\check{s}$ice, Slovak Republic}

\affiliation{$^{3}$Instituto de F\'isica, Universidade Federal de Alagoas, 57072-970 Macei\'o, AL, Brazil}

\begin{abstract}
Phase transitions, compensation phenomenon and magnetization of a ferro-ferrimagnetic ternary alloy AB$_{\rho}$C$_{1-\rho}$ composed of three different kinds of magnetic ions A, B and C  with the spin magnitude 1/2, 1 and 3/2 are examined within the framework of a mixed-spin Ising model on a honeycomb lattice with a selective  annealed site disorder on one of its two sublattices. It is supposed that the first sublattice of a bipartite honeycomb lattice is formed by the spin-1/2 magnetic ions, while the sites of the second sublattice are randomly occupied either by the spin-1 magnetic ions with a probability $\rho$ or the spin-3/2 magnetic ions with a probability $1-\rho$, both being subject to a uniaxial single-ion anisotropy. The model under investigation can be exactly mapped into an effective spin-1/2 Ising model on a triangular lattice through the generalized star-triangle transformation. For a specific concentration of the spin-1 (spin-3/2) magnetic ions, it is shown that the ferro-ferrimagnetic version of the studied model may display a compensation temperature at which the total magnetization vanishes below a critical temperature. The critical temperature strikingly may also become independent of the concentration of the randomly mixed spin-1 and spin-3/2 magnetic ions for a specific value of a uniaxial single-ion anisotropy.
The spontaneous magnetic order may be notably restored at finite temperatures through the order-by-disorder mechanism above a disordered ground state, which results in an anomalous temperature dependence of the total magnetization with double reentrant phase transitions.
\end{abstract}


\maketitle

\section{Introduction}

In the last few decades, advanced magnetic materials with a two-dimensional magnetic structure have attracted a great deal of attention because of their immense technological potential in thermomagnetic recording, electronic, computer technologies, chemical sensors, electronic and optoelectronic devices~\cite{singh,gatteshi,mansuripur,shieh}. Various versions of the mixed-spin Ising models in two dimensions have been extensively studied because they may display richer critical behavior than their single-spin counterparts, along with a greater variability of their magnetic properties (see Ref. \cite{aps} and references cited therein). A considerable attention has been paid, in particular, to the magnetic behavior of ferrimagnetic mixed-spin Ising systems when investigating the role of the spin magnitude and the nature of considered coupling constants upon the magnetic behavior.

Among the most outstanding features of the ferrimagnetic mixed-spin Ising systems belongs the emergence of a compensation phenomenon, which refers to a specific temperature at which the total magnetization of a magnetic system vanishes in spite of an existent spontaneous long-range magnetic order at a given temperature. The appearance of a compensation point is closely related to a complete cancellation of the magnetic moments of at least two different magnetic species (magnetic ions), which are coupled through the antiferromagnetic interaction and may naturally display different temperature dependencies of the relevant sublattice magnetizations. It should be stressed, moreover, that the compensation phenomenon is of particular technological relevance for thermomagnetic recording, because only a small change of a driving field is needed for a reversal of the total magnetization \cite{mansuripur,shieh}.

The binary mixed-spin Ising systems are the simplest models allowing a theoretical description of the main characteristics of ferrimagnetism. To date, there are numerous theoretical studies of the binary mixed-spin Ising models defined on a honeycomb lattice \cite{strecka,ekiz,jascur,deviren,jurcisin,dakhama,nakamura,figueiredo,godoy,keskin,gomez,albarracin}, bathroom-tile lattice \cite{str06,masja}, diced lattice \cite{stre}, square lattice \cite{buendia,kaneyoshi,mohamad,masrour,bouda}, Bethe lattice \cite{albayrak,karimou}, etc. As far as spin sizes are concerned, the binary mixed-spin Ising models with a great variety of combinations of the spin magnitudes have been investigated as, for instance, (1/2, 1)~\cite{strecka,figueiredo,ekiz,godoy,stre}, (1/2, 3/2)~\cite{jascur,buendia}, (1, 3/2)~\cite{deviren,jurcisin}, (1/2, $S>1/2$)~\cite{kaneyoshi,dakhama,str06}, (2, 5/2)~\cite{nakamura,keskin}, (3/2, 2)~\cite{albayrak}, (7/2, 1)~\cite{karimou,mohamad}, (7/2, 2)~\cite{masrour} and (7/2, 3)~\cite{bouda}. Magnetic properties of the binary mixed-spin Ising models on a honeycomb lattice were exploited by various theoretical methods such as exact mapping technique \cite{strecka,jascur}, Monte Carlo simulations~\cite{figueiredo,nakamura,buendia,gomez}, cluster variational method~\cite{tucker}, effective- \cite{deviren} and mean-field \cite{kaneyoshi1} theories.

Compared to the above, the ternary mixed-spin Ising systems built out of three different magnetic species were up to now much less studied and their magnetic characteristics remain far from being fully understood yet.  So far, there are only a few rigorous studies of the ternary mixed-spin Ising models on decorated square \cite{scj07,csj08} and Bethe \cite{dev08,dev09,dev10} lattices. Even more complicated situation emerges when considering the mixed-spin Ising models for ternary alloys, which are being subject to a site disorder emergent leastwise at one of its sublattices. This situation is experimentally encountered, for instance, in Prussian blue analogs (C$_{1-\rho}$Mn$_{\rho}$)$_3$[Cr(CN)$_6$]$_2$ $\cdot$ n H$_2$O (C = Ni$^{2+}$ or Fe$^{2+}$), which provide experimental realizations of the ternary alloy AB$_{\rho}$C$_{1-\rho}$ with a perfect occupation of the first sublattice by the spin-3/2 magnetic ions Cr$^{3+}$ and a site randomness of the second sublattice occupied with a probability $\rho$ by the spin-5/2 magnetic ions Mn$^{2+}$and with a probability $1-{\rho}$ either by the spin-1 magnetic ions Ni$^{2+}$ or the spin-2 magnetic ions Fe$^{2+}$, respectively \cite{ohk97a,ohk97b, ohk97c}. The phase diagrams, compensation phenomenon, and magnetic properties of the mixed-spin Ising ternary alloys AB$_{\rho}$C$_{1-\rho}$ with various combinations of the spin magnitudes were thoroughly described by Bob\'ak and coworkers by making use of the mean-field theory \cite{bob03,bob04,bob06,bob07}, effective-field theory \cite{bob02} and Monte Carlo simulations \cite{gbuendia,bob09,bob10}. To the best of our knowledge, the exact calculations for the mixed-spin Ising ternary alloys AB$_{\rho}$C$_{1-\rho}$ have not been reported in the literature yet.

In the present work we will therefore introduce and exactly solve the mixed spin-1/2, 1 and 3/2 Ising model on a bipartite honeycomb lattice, which will be designed for a ternary alloy AB$_{\rho}$C$_{1-\rho}$ with a selective site disorder on one of its two sublattices. In the section II we will at first describe the investigated model and basic steps of its exact solution following the approach elaborated previously for the mixed-spin Ising model on a selectively diluted honeycomb lattice \cite{aps}. The section III elucidates phase diagrams, critical behavior and compensation phenomenon of the studied system. Typical temperature variations of the spontaneous magnetization will be reported in the section IV. Finally, some conclusions and future outlooks are mentioned in the section V.

\section{Model and its exact solution}

\begin{figure}[h]
\includegraphics[scale=1]{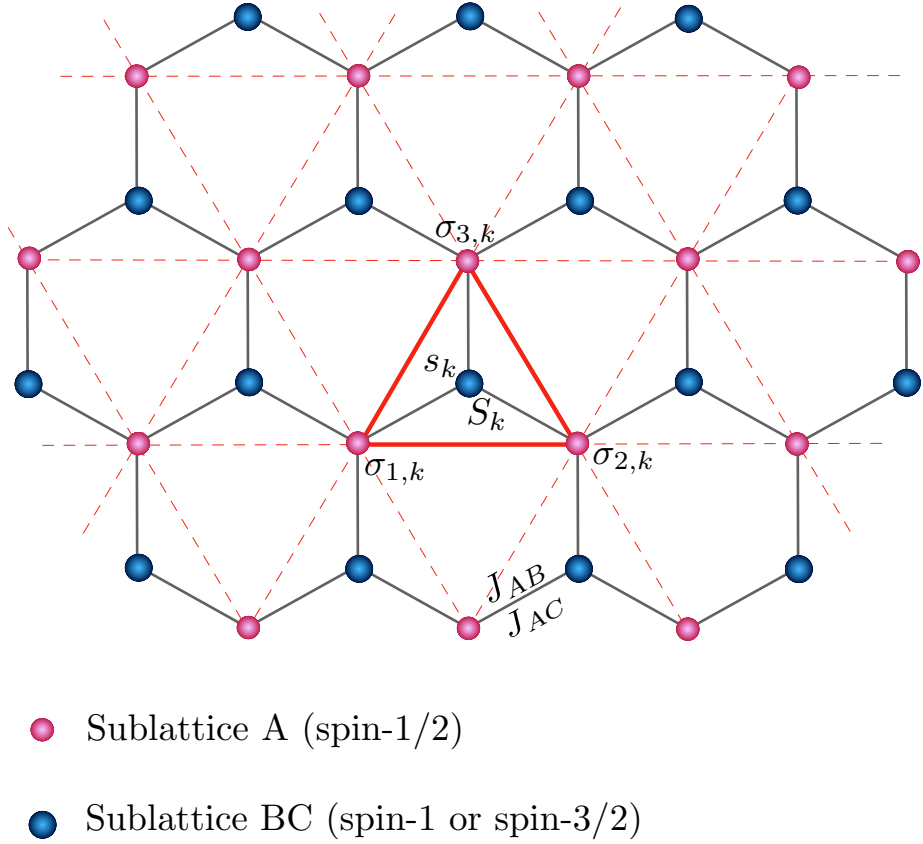} \caption{\label{fig:1} (Color online) A schematic representation of a honeycomb lattice. Small circles represent the spin-1/2 magnetic ions ($\sigma$) from the sublattice A, while the large circles denote the randomly mixed spin-1 ($s$) and spin-3/2 ($S$) magnetic ions from the sublattice BC. Solid lines stand for the coupling constant $J_{\mathrm{AB}}$ or $J_{\mathrm{AC}}$, while broken lines visualize effective couplings within a triangular lattice obtained after a star-triangle transformation.}
\end{figure}

Let us begin by considering the mixed spin-1/2, spin-1 and spin-3/2 Ising model for the ternary alloy with a selective annealed site disorder  on a given sublattice of a honeycomb lattice as depicted in Figure~\ref{fig:1}. The small circles ($\sigma$) represent the spin-1/2 magnetic ions forming the sublattice A and the large circles correspond either to the spin $s=1$ or $S=3/2$ magnetic ions B or C forming the other sublattice BC. It is worthy to mention that the magnetic ions B and C are both randomly distributed over the same sublattice BC.
Then, the total Hamiltonian can be written as:
\begin{align}
\mathcal{H}=-J_{\mathrm{AB}}\sum_{<i,j>} \sigma_i s_j n_j -J_{\mathrm{AC}}\sum_{<i,j>} \sigma_i S_j (1-n_j)
-D_{\mathrm{B}}\sum_{j=1}^Ns_j^2n_j -D_{\mathrm{C}}\sum_{j=1}^NS_j^2(1-n_j),
\label{eq1}
\end{align}
where the summation symbol $<i,j>$ runs over all nearest-neighbor spin pairs, the index $i$ refers to a lattice site from the sublattice A, the index $j$ corresponds to a lattice site from the sublattice BC with the spin size $s_j=1$ or $S_j=3/2$, respectively, and $N$ denotes the total number of the lattice sites within the sublattice A. The coupling constant $J_{\mathrm{AB}}$ denotes the nearest-neighbor interaction between the spin-1/2 and spin-1 magnetic ions, while the coupling constant $J_{\mathrm{AC}}$ stands for the nearest-neighbor interaction between the spin-1/2 and spin-3/2 magnetic ions. The parameters $D_{\mathrm{B}}$ and $D_{\mathrm{C}}$ measure a strength of the uniaxial single-ion anisotropy acting on the spin-1 and spin-3/2 magnetic ions from the sublattice BC. Finally, the two-valued random variable $n_j=0,1$ determines whether the lattice site $j$ in the sublattice BC is occupied by the spin-1 magnetic ion  B ($n_j=1$) or by the spin-3/2 magnetic ion C ($n_j=0$). The distribution function for the binary random variable $n_j$ is given by:
\begin{align}
P(n_j) = \rho \delta(n_j-1) + (1-\rho) \delta(n_j),
\label{df}
\end{align}
where the probability $\rho$ determines a concentration of the spin-1  magnetic ions  B and the probability $1-\rho$ determines a concentration of the spin-3/2 magnetic ions C.

The mixed-spin Ising model with a selective site disorder on a honeycomb lattice can be rigorously solved following the approach elaborated previously for the mixed-spin Ising model on a selectively diluted honeycomb lattice \cite{aps}. Let us rewrite first the Hamiltonian (\ref{eq1}) as a sum of the cluster Hamiltonians $\mathcal{H}=\sum_{k}\mathcal{H}_k$, each of which involves all interaction terms pertinent to a single four-spin cluster with the geometric shape of a star (see the four-spin cluster delimited  in Fig. \ref{fig:1} by a thick red triangle):
\begin{align}
\mathcal{H}_k=-J_{\mathrm{AB}}s_k n_k(\sigma_{1,k}+\sigma_{2,k}+\sigma_{3,k})-J_{\mathrm{AC}}S_k (1-n_k)(\sigma_{1,k}+\sigma_{2,k}+\sigma_{3,k}) -D_{\mathrm{B}}s_k^2n_k -D_{\mathrm{C}}S_k^2(1-n_k).
\label{eq2}
\end{align}
The relevant Boltzmann's factor corresponding to the cluster Hamiltonian (\ref{eq2}) reads as follows:
\begin{align}
\omega(\{\sigma_k\})=\sum_{S_k=-\frac{3}{2}}^{\frac{3}{2}}\sum_{s_k=-1}^{1} \sum_{n_k=0}^{1} \mathrm{e}^{-\beta \mathcal{H}_k +\beta \mu n_k},
\label{eq4}
\end{align}
where $\{\sigma_k\}\equiv \sigma_{1,k}+\sigma_{2,k}+\sigma_{3,k}$, $\beta=1/(k_BT)$, $k_B$ is the Boltzmann constant, $T$ denotes the absolute temperature and $\mu$ is the chemical potential of the spin-1 magnetic ions B. Within this approach, we are assuming an equilibrium disorder distribution, usually termed as annealed disorder. In systems with quenched disorder the distribution is not thermalized. Due to the absence of spin frustration in this lattice, one expects the thermodynamic magnetic behavior to be quite similar irrespective to the quenched or annealed nature of the disorder distribution. For the sake of brevity, let us introduce the fugacity of the spin-1 magnetic ions $z=$e$^{\beta \mu}$ and the parameter $\zeta_k=\sigma_{1,k}+\sigma_{2,k}+\sigma_{3,k}$, which enable us to rewrite the Boltzmann's factor \eqref{eq4} into the following explicit form after performing all three summations in Eq. \eqref{eq4}:
\begin{eqnarray}
\omega(\zeta_k)=&z+2z\mathrm{e}^{\beta D_{\mathrm{B}}}\cosh\left(\beta J_{\mathrm{AB}}\zeta_k\right)
+2\mathrm{e}^{\frac{9\beta D_{\mathrm{C}}}{4}}\cosh\left(\frac{3\beta J_{\mathrm{AC}}}{2}\zeta_k\right)+2\mathrm{e}^{\frac{\beta D_{\mathrm{C}}}{4}}\cosh\left(\frac{\beta J_{\mathrm{AC}}}{2}\zeta_k\right).
\label{eq5}
\end{eqnarray}
The generalized star-triangle transformation \cite{Fisher} allows one to replace the Boltzmann factor \eqref{eq5} through the equivalent expression given by the effective Boltzmann weight
\begin{eqnarray}
 \tilde{\omega}(\zeta_k)=
 A\mathrm{e}^{\beta R(\sigma_{1,k}\sigma_{2,k}+\sigma_{2,k}\sigma_{3,k}+\sigma_{3,k}\sigma_{1,k})}.
\label{bwe}
\end{eqnarray}
So far not specified mapping parameters $A$ and $R$ can be simply obtained by imposing the self-consistency condition of the star-triangle transformation $\omega(\zeta_k)=\tilde{\omega}(\zeta_k)$, which necessitates equality between the Boltzmann weights \eqref{eq5} and \eqref{bwe} for all available combinations of the spin values $\sigma_{1,k}$, $\sigma_{2,k}$ and $\sigma_{3,k}$. One may easily convince oneself that the star-triangle transformation $\omega(\zeta_k)=\tilde{\omega}(\zeta_k)$ is in fact just a set of two independent equations :
\begin{eqnarray}
\label{omega1}
  \omega\left(\tfrac{3}{2}\right)=&zV_1+V_3=\tilde{\omega}\left(\tfrac{3}{2}\right)=A\mathrm{e}^{\frac{3\beta R}{4}},\\
  \omega\left(\tfrac{1}{2}\right)=&
 zV_2+V_4=\tilde{\omega}\left(\tfrac{1}{2}\right)=
  A\mathrm{e}^{-\frac{\beta R}{4}}.\label{omega2}
\end{eqnarray}
In above, we have defined the following four functions:
\begin{align}
V_1 =& 1+2\mathrm{e}^{\beta D_{\mathrm{B}}}\cosh\left(\frac{3\beta J_{\mathrm{AB}}}{2}\right),\label{v1}\\
V_2 =&1+2\mathrm{e}^{\beta D_{\mathrm{B}}}\cosh\left(\frac{\beta J_{\mathrm{AB}}}{2}\right),\label{v2}\\
V_3 =&2\mathrm{e}^{\frac{9\beta D_{\mathrm{C}}}{4}}\cosh\left(\frac{9\beta J_{\mathrm{AC}}}{4}\right)+2e^{\frac{\beta D_{\mathrm{C}}}{4}}\cosh\left(\frac{3\beta J_{\mathrm{AC}}}{4}\right),\label{v3}\\
V_4 =&2\mathrm{e}^{\frac{9\beta D_{\mathrm{C}}}{4}}\cosh\left(\frac{3\beta J_{\mathrm{AC}}}{4}\right)+2e^{\frac{\beta D_{\mathrm{C}}}{4}}\cosh\left(\frac{\beta J_{\mathrm{AC}}}{4}\right)\label{v4}.
\end{align}
By solving the couple of Eqs.~\eqref{omega1} and \eqref{omega2} one gets the following exact formula for the mapping parameter
\begin{align}
A=\left[\left(zV_1+V_3\right)\left(zV_2+V_4\right)^3\right]^{\frac{1}{4}}
\label{eq7}
\end{align}
and
\begin{align}
R= \frac{1}{\beta}\ln\left(\frac{zV_1+V_3}{zV_2+V_4}\right).
\label{eq6}
\end{align}
After plugging in the star-triangle transformation into the grand-canonical partition function of the mixed spin-1/2, 1 and 3/2 Ising ternary alloy on a selectively disordered honeycomb lattice one obtains the exact mapping relationship with the partition function of the simple spin-1/2 Ising model on a triangular lattice:
\begin{align}
\Xi=\sum_{\{\sigma_k\}}\prod_{k=1}^N \omega(\{\sigma_k\})=
    \sum_{\{\sigma_k\}}\prod_{k=1}^N \tilde{\omega}(\{\sigma_k\})= A^N \mathcal{Z}_t(\beta , R),
\label{eq3}
\end{align}
which is defined through the effective Hamiltonian with temperature-dependent nearest-neighbor interaction $R$:
\begin{eqnarray}
  \mathcal{H}_t= - R \sum_{<i,j>} \sigma_i\sigma_j.
\label{hame}	
\end{eqnarray}
The grand potential of the mixed-spin Ising ternary alloy on a selectively disordered honeycomb lattice consequently reads:
\begin{align}
\Omega = -k_B T \ln \Xi = -N k_B T \ln A - k_B T \ln \mathcal{Z}_t(\beta , R).
\label{eq14}
\end{align}
From the above equation one may simply calculate the mean value for the concentration of the spin-1 magnetic ions using the formula:
\begin{align}
\rho = -\frac{1}{N}\frac{\partial \Omega}{\partial \mu}
     = \frac{z\beta}{N}\frac{\partial \Omega}{\partial z}
     = z\frac{\partial \ln A}{\partial z}+3\beta\varepsilon\frac{\partial R}{\partial z},
\label{eq13}
\end{align}
where $\varepsilon=\langle\sigma_i\sigma_j\rangle_t$ denotes the nearest-neighbor pair correlation function of the effective spin-1/2 Ising model on a triangular lattice given by the effective Hamiltonian \eqref{hame}. After some algebraic manipulations, one may express the concentration $\rho$ of the spin-1 magnetic ions in the following compact form:
\begin{align}
\rho = \frac{zV_1}{zV_1+V_3}\left(\frac{1}{4}+3\varepsilon\right)+\frac{3zV_2}{zV_2+V_4}\left(\frac{1}{4}-\varepsilon\right).
\label{eq15}
\end{align}
The formula \eqref{eq15} represents a central result of our calculation as it can be viewed as an equation of state, from which phase diagrams, critical behavior as well as all basic thermodynamic quantities can be rigorously calculated when the concentration of the spin-1 magnetic ions is fixed to some specific value. It is worthwhile to recall that the particular case $\rho=1$ corresponds to a mixed spin-1/2 and spin-1 Ising model on a honeycomb lattice \cite{strecka}, while the other special case $\rho=0$ corresponds to a mixed spin-1/2 and spin-3/2 Ising model on a honeycomb lattice \cite{jascur}. The selective site randomness can be thus examined within a range of the concentration $0<\rho<1$, whereas the strongest effect of a selective site disorder could be expected for the concentration $\rho = 0.5$ assuming half of the lattice sites occupied by the spin-1 magnetic ions and another half of the lattice sites occupied by the spin-3/2 magnetic ions.

\section{Phase transitions and critical phenomena}

In this section we will establish finite-temperature phase diagrams of the mixed-spin Ising ternary alloy on a selectively diluted honeycomb lattice. To this end, it is sufficient to realize that the mapping relation \eqref{eq3} between the grand-canonical partition function of the mixed-spin Ising ternary alloy on a selectively diluted honeycomb lattice shows a singularity only if the same singularity appears in the canonical partition function of the effective spin-1/2 Ising model on a triangular lattice given by the Hamiltonian \eqref{hame}. It should be pointed out that the critical parameters for the spin-1/2 Ising model on a triangular lattice are known exactly (for  review see for instance Refs. \cite{aps,domb}): the inverse critical temperature reads $\beta_c R = R/(k_B T_c) = \ln 3$ and the critical value of the nearest-neighbor pair correlation function is $\varepsilon_c=\frac{1}{6}$. The critical value of the fugacity $z_c$ can be consequently obtained from Eq.~\eqref{eq6}:
\begin{eqnarray}
  \beta_c R = \ln \left(\frac{z_cV_1^c+V_3^c}{z_cV_2^c+V_4^c}\right)
	\quad \Rightarrow \quad z_c=\frac{V_3^c-3V_4^c}{3V_2^c-V_1^c}.
  \label{zcrit}
\end{eqnarray}
In above, the superscript $c$ at the relevant expressions $V_j^c (j=1-4)$ means that the inverse critical temperature $\beta_c = 1/(k_BT_c)$ enters into their definitions \eqref{v1}-\eqref{v4} instead of $\beta$. Now, one may substitute the critical value of the fugacity $z_c$ given by Eq. \eqref{zcrit} and the nearest-neighbor pair correlation function  $\varepsilon_c=1/6$ into Eq.~\eqref{eq15} in order to get the critical condition of the mixed-spin Ising model on a selectively disordered honeycomb lattice:
\begin{align}
\rho_c=\frac{\left(V_1^c+V_2^c\right)\left(3V_4^c-V_3^c\right)}{4\left(V_1^cV_4^c-V_2^cV_3^c\right)}.
\label{eqcc}
\end{align}
The critical boundaries between the spontaneously ordered and disordered phases of the mixed-spin Ising model on a selectively disordered honeycomb lattice can be now straightforwardly obtained by solving numerically the critical condition \eqref{eqcc}. Before presenting a few typical finite-temperature phase diagrams, it is worthwhile to remark that both considered coupling constants $J_{\mathrm{AB}}$ and $J_{\mathrm{AC}}$ enter in the expressions $V_j^c (j=1-4)$ given by Eqs. \eqref{v1}-\eqref{v4} just in the arguments of even functions, which means that a relative size of the critical temperature is independent of whether these coupling constants are being considered ferromagnetic ($J_{AB}, J_{AC}>0$) or antiferromagnetic ($J_{\mathrm{AB}}, J_{\mathrm{AC}}<0$). In what follows all the interaction terms will be accordingly scaled with respect to a size of the coupling constant $|J_{AB}|$, whereas both parameters of the uniaxial single-ion anisotropy will be set equal to each other $D_{\mathrm{B}}=D_{\mathrm{C}}=D$.

\begin{figure}
\includegraphics[scale=0.5]{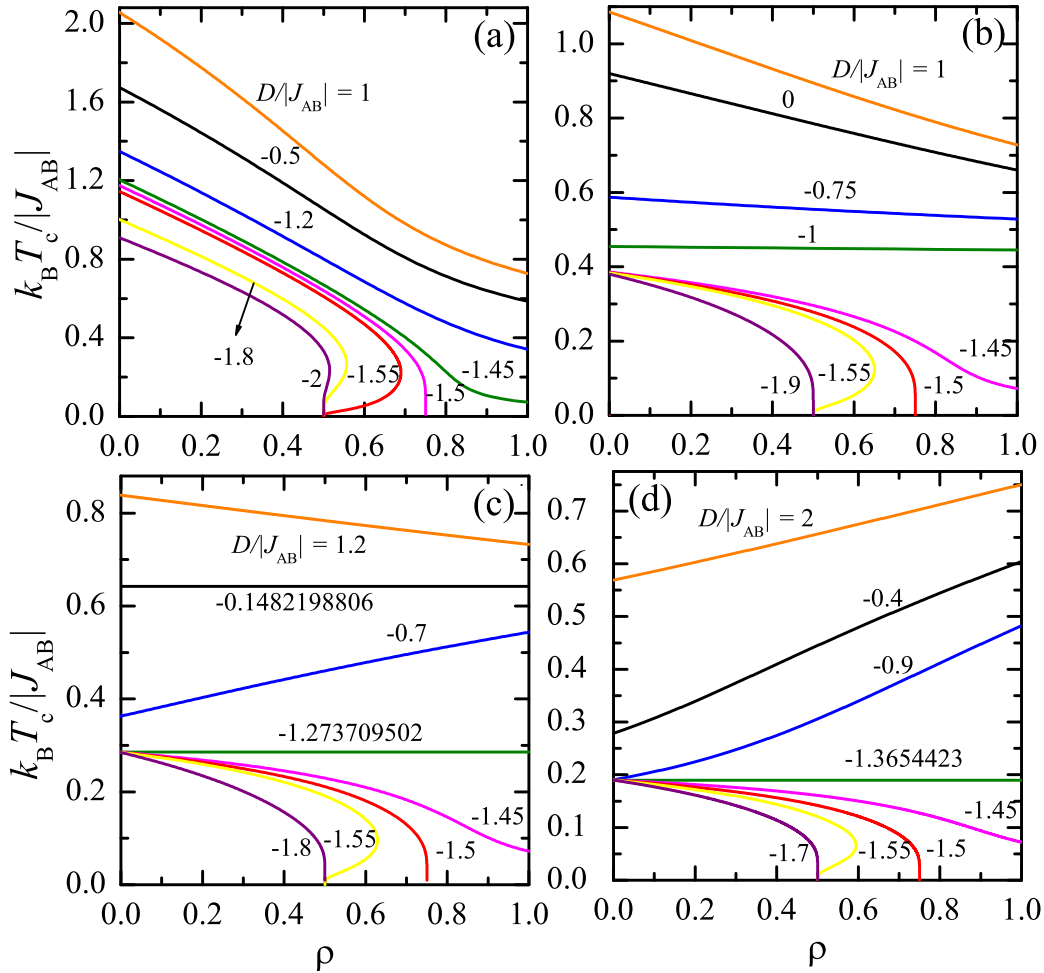}
\caption{\label{fig:2} Critical temperature $k_BT_c/|J_{\mathrm{AB}}|$ as a function of the concentration of the spin-1 magnetic ions $\rho$ for several values of the uniaxial single-ion anisotropy $D/|J_{\mathrm{AB}}|$ and four different values of the interaction ratio: (a) $|J_{\mathrm{AC}}|/|J_{\mathrm{AB}}|=2$; (b) $|J_{\mathrm{AC}}|/|J_{\mathrm{AB}}|=1$; (c) $|J_{\mathrm{AC}}|/|J_{\mathrm{AB}}|=0.75$;
(d) $|J_{\mathrm{AC}}|/|J_{\mathrm{AB}}|=0.5$.}
\end{figure}

The reduced critical temperature $k_BT_c/|J_{\mathrm{AB}}|$ is plotted in Fig.~\ref{fig:2}(a)-(d) against the concentration $\rho$ for several values of the uniaxial single-ion anisotropy $D/|J_{\mathrm{AB}}|$. It is quite evident from Fig.~\ref{fig:2}(a)-(d) that the investigated mixed-spin system displays at sufficiently low temperatures a spontaneous magnetic ordering regardless of the concentration $\rho$ for any $D/|J_{\mathrm{AB}}|>-1.5$, while the ground state becomes disordered at low enough temperatures for any $D/|J_{\mathrm{AB}}|<-1.5$ provided that high concentrations $\rho>0.5$ of the spin-1 magnetic ions are assumed. This latter finding can be attributed to the effect of nonmagnetic dilution \cite{aps}, because the spin-1 magnetic ions preferentially occupying the selectively diluted sublattice for the concentrations $\rho>0.5$ are forced by strong enough uniaxial single-ion anisotropy $D/|J_{AB}|<-1.5$ to their nonmagnetic state $s_j=0$. The most remarkable finding is, however, that the spontaneous magnetic ordering can be restored at finite temperatures also in the parameter region $\rho \gtrsim 0.5$ and $D/|J_{\mathrm{AB}}|\lesssim-1.5$ above the disordered ground state through the order-by-disorder mechanism \cite{vil80,she82}. It actually turns out that the investigated mixed-spin system shows, in this parameter space, double reentrant phase transitions. This phenomenon occurs whenever the system orders above a disordered ground state at a lower critical temperature and this spontaneous magnetic long-range order persists up to higher critical temperature. Last but not least, it is worth mentioning that the critical temperature of the mixed-spin Ising model on a selectively disordered honeycomb lattice  may strikingly remain constant upon change of the concentration $\rho$ for specific values of the uniaxial single-ion anisotropy as, for instance, $D/|J_{\mathrm{AB}}|=-1$ for $|J_{\mathrm{AC}}|/|J_{\mathrm{AB}}|=1$ [Fig.~\ref{fig:2}(b)], $D/|J_{\mathrm{AB}}|\approx-0.15$ or $-1.275$ for $|J_{\mathrm{AC}}|/|J_{\mathrm{AB}}|=0.75$ [Fig.~\ref{fig:2}(c)] or
$D/|J_{\mathrm{AB}}|\approx-1.365$ for $|J_{\mathrm{AC}}|/|J_{\mathrm{AB}}|=0.5$ [Fig.~\ref{fig:2}(d)].

\begin{figure}
\includegraphics[scale=0.5]{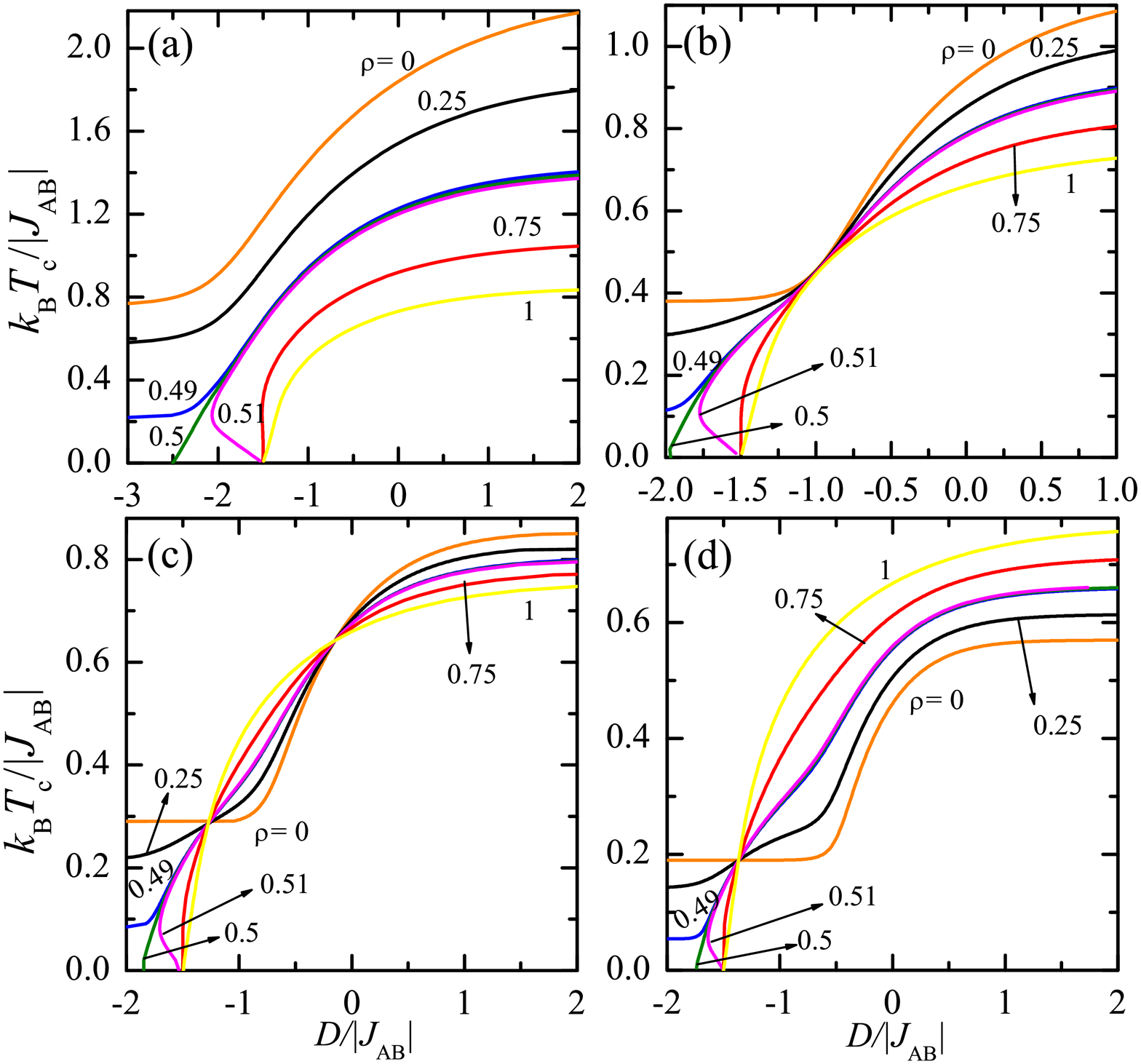}
\caption{\label{fig:3} Critical temperature $k_BT_c/|J_{\mathrm{AB}}|$ as a function of the uniaxial single-ion anisotropy $D/|J_{\mathrm{AB}}|$ for several values of the concentration $\rho$ and four different values of the interaction ratio: (a) $|J_{\mathrm{AC}}|/|J_{\mathrm{AB}}|=2$; (b) $|J_{\mathrm{AC}}|/|J_{\mathrm{AB}}|=1$; (c) $|J_{\mathrm{AC}}|/|J_{\mathrm{AB}}|=0.75$;
(d) $|J_{\mathrm{AC}}|/|J_{\mathrm{AB}}|=0.5$.}
\end{figure}

Next, the critical temperature $k_BT_c/|J_{\mathrm{AB}}|$ is displayed in Fig. \ref{fig:3}(a)-(d) as a function of the uniaxial single-ion anisotropy $D/|J_{\mathrm{AB}}|$ for several values of the concentration $\rho$. Note that the dependencies of the critical temperature on the uniaxial single-ion anisotropy for the particular values of the concentration sufficiently close to $\rho = 1$ ($\rho = 0$) are quite typical for the mixed spin-1/2 and spin-1 (spin-3/2) Ising models on a honeycomb lattice \cite{strecka,jascur}. It is also obvious from Fig. \ref{fig:3}(a)-(d) that the phase boundaries for the concentration $\rho = 0.51$ verify existence of double reentrant phase transitions in the parameter region $D/|J_{\mathrm{AB}}|\lesssim-1.5$ and $\rho \gtrsim 0.5$ with the disordered ground state. Moreover, the critical boundaries depicted in Fig.~\ref{fig:3}(b)-(d) provide an alternative confirmation of the independence of the critical temperature on the concentration $\rho$ when all displayed critical frontiers for the interaction ratio $|J_{\mathrm{AC}}|/|J_{\mathrm{AB}}|\leq 1$ intersect each other either in one [Fig.~\ref{fig:3}(b),(d)] or two [Fig.~\ref{fig:3}(c)] common points. Our exact results thus clearly corroborate independence of the critical temperature upon change of the concentration of the mixed-spin Ising model on a selectively disordered honeycomb lattice, which has been reported by the first time by Bob\'ak and coworkers using the approximate mean-field and effective-field theories \cite{bob02,bob03,bob04}.

\begin{figure}
\includegraphics[scale=0.5]{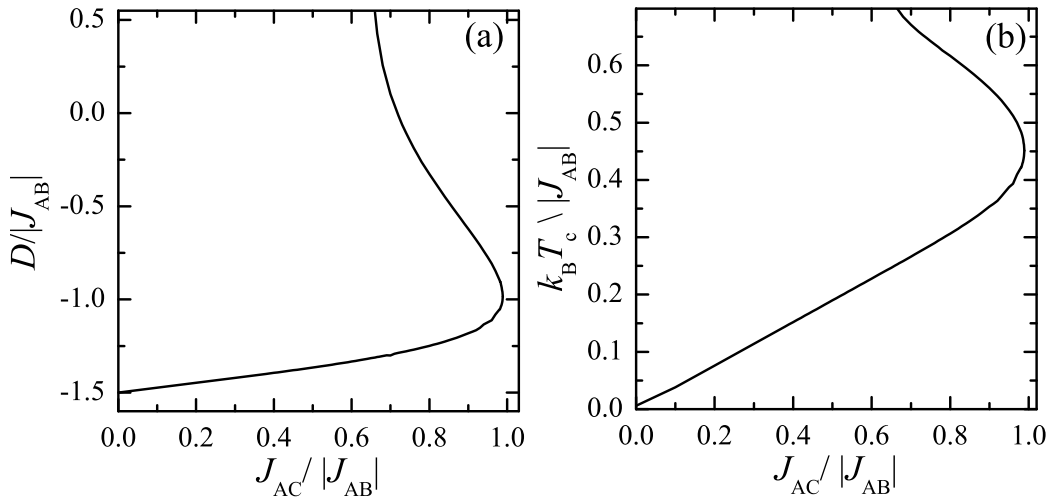}
\caption{\label{fig:4} (a) The single-ion anisotropy $D/|J_{\mathrm{AB}}|$ versus the interaction ratio $|J_{\mathrm{AC}}|/|J_{\mathrm{AB}}| $ dependence along which the critical temperature is kept constant regardless of the concentration  of the spin-1 magnetic ions $\rho$;
(b) The corresponding critical temperature $k_BT_c/|J_{\mathrm{AB}}|$ versus the interaction ratio $|J_{\mathrm{AC}}|/|J_{\mathrm{AB}}| $ dependence.}
\end{figure}

It is clear that this latter finding could be of considerable technological relevance, because the change in a chemical composition (i.e. the concentration of the magnetic ions) would not affect under this specific constraint temperature range for applicability of a given magnetic material. For this reason, we have decided to illustrate in Fig. \ref{fig:4}(a) the dependence of the single-ion anisotropy $D/|J_{\mathrm{AB}}|$ on the interaction ratio $|J_{\mathrm{AC}}|/|J_{\mathrm{AB}}|$ along which the critical temperature is kept constant regardless of the concentration $\rho$. According to this plot, the mixed-spin Ising ternary alloy may indeed display this intriguing feature either for one or two specific values of the uniaxial single-ion anisotropy $D/|J_{\mathrm{AB}}|$ whenever the interaction ratio $|J_{\mathrm{AC}}|/|J_{\mathrm{AB}}|\leq 1$. The isotropic interaction ratio $|J_{\mathrm{AC}}|/|J_{\mathrm{AB}}|=1$ accordingly provides an upper limit for an observation of this striking phenomenon, whereas in this special case all critical lines touch rather than cross each other [c.f. Fig.~\ref{fig:3}(b)]. A relative size of the critical temperature for those special cases  is depicted in Fig. \ref{fig:4}(b).

\section{Magnetization and compensation phenomenon}

In this section we will turn our attention to a detailed investigation of temperature dependence of the spontaneous magnetization, which will serve as the order parameter characterizing the nature of a spontaneous long-range magnetic ordering. Of course, three different sublattice magnetizations have to be computed first in order to get the total spontaneous magnetization. It can be readily proved by exploiting exact mapping theorems developed by Barry \textit{et al}. \cite{bar88,bar90,bar91,bar95} that the spontaneous magnetization of the sublattice A formed by the spin-1/2 magnetic ions directly equals to the spontaneous magnetization of the effective spin-1/2 Ising model on a triangular lattice:
\begin{eqnarray}
m_{\mathrm{A}}=\langle\sigma_k^z\rangle=\langle\sigma_k^z\rangle_{t}=m_{t}(\beta R),
\end{eqnarray}
which has been exactly calculated by Potts \cite{potts}:
\begin{eqnarray}
m_{t}=\frac{1}{2}\left[1-\frac{16y^6}{(1+3y^2)(1-y^2)^3}\right]^{\frac{1}{8}}, \qquad y=\exp(-\beta R/2).
\label{mit}
\end{eqnarray}
In this regard, an exact expression for the spontaneous magnetization of the sublattice A formed by the spin-1/2 magnetic ions readily follows from Eq. \eqref{mit} when substituting therein the explicit form of effective nearest-neighbor coupling \eqref{eq6}:
\begin{eqnarray}
m_{\mathrm{A}}=\frac{1}{2}\left[1-\frac{16(zV_1+V_3)(zV_2+V_4)^3} {\left(zV_1+V_3+3zV_2+3V_4\right)\left(zV_1+V_3-zV_2-V_4\right)^3} \right]^\frac{1}{8}.
\label{eq16}
\end{eqnarray}

On the other hand, an exact calculation of the spontaneous magnetization of the sublattice B and C formed by the spin-1 and spin-3/2 magnetic ions can be performed with the help of the generalized Callen-Suzuki identity \cite{callen,Suzuki,balcerzak}. For instance, the spontaneous magnetization of the sublattice B composed of the spin-1 magnetic ions follows from the exact spin identity:
\begin{eqnarray}
  m_{\mathrm{B}} \equiv \langle n_j \langle s_k\rangle \rangle_c = \rho \left\langle\frac{\displaystyle\sum_{ s_k=-1}^1s_k\mathrm{e}^{-\beta \mathcal{H}_k}}{\displaystyle\sum_{s_k=-1}^1\mathrm{e}^{-\beta \mathcal{H}_k}}\right\rangle= \rho \left\langle\frac{2\mathrm{e}^{\beta D_{\mathrm{B}}}\sinh\left(\beta J_{\mathrm{AB}}\zeta_k\right)}{1+2\mathrm{e}^{\beta D_{\mathrm{B}}}\cosh\left(\beta J_{\mathrm{AB}}\zeta_k\right)}\right\rangle,
  \label{ms}
\end{eqnarray}
where the symbols $\langle \cdots \rangle_c$ and $\langle \cdots \rangle$ denote a configurational average and canonical ensemble average, respectively. To obtain the statistical mean value on the right-hand-side of Eq. \eqref{ms} one may perform the following expansion:
\begin{eqnarray}
  \langle f\left(\sigma_{1,k},\sigma_{2,k},\sigma_{3,k}\right)\rangle=
  c_1+c_2\langle\left(\sigma_{1,k}+\sigma_{2,k}+
  \sigma_{3,k}\right)\rangle+
  c_3\langle\left(\sigma_{1,k}\sigma_{2,k}+
  \sigma_{2,k}\sigma_{3,k}
  +\sigma_{3,k}\sigma_{1,k}\right)\rangle+c_4\langle\left(\sigma_{1,k}
  \sigma_{2,k}\sigma_{3,k}\right)\rangle,
\end{eqnarray}
which simplifies in a zero magnetic field to the following form due to zero value of the coefficients $c_1$ and $c_3$:
\begin{eqnarray}
   \langle f\left(\sigma_{1,k},\sigma_{2,k},\sigma_{3,k}\right)\rangle=
   c_2\langle\left(\sigma_{1,k}+\sigma_{2,k}+\sigma_{3,k}\right)\rangle
   +c_4\langle\sigma_{1,k}\sigma_{2,k}\sigma_{3,k}\rangle = 3 c_2 m_{\mathrm{A}} + c_4 t_{\mathrm{A}}.
   \label{fmag}
\end{eqnarray}
Besides the spontaneous magnetization of the sublattice A given by Eq. \eqref{eq16} one should also calculate the triplet correlation function $t_{\mathrm{A}} \equiv \langle\sigma_{1,k}\sigma_{2,k}\sigma_{3,k}\rangle$  in order to complete an exact calculation of the spontaneous magnetization of the sublattice B. To this end, one may use the exact result for the triplet correlation function of the effective spin-1/2 Ising model on a triangular lattice derived by Baxter and Choy \cite{baxter}:
\begin{align}
t_{\mathrm{A}}=\frac{m_{\mathrm{A}}}{4}\left[1-2\frac{3x^2- 3y^2-2x\sqrt{x^2+2xy-3y^2}}{(x-y)^2}\right], \label{eq26}
\end{align}
with $x=$e$^{\beta R}$ and $y=$e$^{-\beta R/2}$. A substitution of the effective nearest-neighbor coupling \eqref{eq6} into Eq. \eqref{eq26} gives the following exact expression for the triplet correlation function:
\begin{align}
t_{\mathrm{A}}=\frac{m_{\mathrm{A}}}{4}\left[\frac{3(zV_1+V_3)^2-3(zV_2+V_4)^2-2(zV_1+V_3)
\sqrt{\left(z(V_1+V_2)+V_3+V_4\right)^2-4(zV_2+V_4)^2}}
{(zV_1+V_3-zV_2-V_4)^2}\right]. \label{eq27}
\end{align}
The spontaneous magnetization of the  spin-1 magnetic ions B can be now calculated according to Eq. \eqref{fmag}:
\begin{eqnarray}
  m_{\mathrm{B}} = \frac{3}{2}\rho m_{\mathrm{A}}\left[f\left(\tfrac{3}{2}\right)+ f\left(\tfrac{1}{2}\right)\right]+ 2 \rho t_{\mathrm{A}}\left[f\left(\tfrac{3}{2}\right) -3f\left(\tfrac{1}{2}\right)\right],
\end{eqnarray}
where the explicit form of the function $f(x)$ reads as follows:
\begin{eqnarray}
  f(x)= \frac{2\mathrm{e}^{\beta D_{\mathrm{B}}}\sinh\left(\beta J_{\mathrm{AB}}x\right)}{1+2\mathrm{e}^{\beta D_{\mathrm{B}}}\cosh\left(\beta J_{\mathrm{AB}}x\right)}.
  \label{fx}
\end{eqnarray}
The spontaneous magnetization of the ions B can be alternatively expressed in terms of the parameters $V_1$ and $V_2$ given by Eqs. \eqref{v1}-\eqref{v2}:
\begin{eqnarray}
m_{\mathrm{B}}=\frac{3}{2} \rho  m_{\mathrm{A}} \left(\frac{W_1}{V_1}+\frac{W_2}{V_2}\right)+2 \rho t_{\mathrm{A}}\left(\frac{W_1}{V_1}-\frac{3W_2}{V_2}\right)
\label{eq17}
\end{eqnarray}
and two newly defined functions:
\begin{align}
W_1=&2e^{\beta D_{\mathrm{B}}}\sinh\left(\frac{3\beta J_{\mathrm{AB}}}{2}\right),\label{eq19} \\
W_2=&2e^{\beta D_{\mathrm{B}}}\sinh\left(\frac{\beta J_{\mathrm{AB}}}{2}\right). \label{eq20}
\end{align}

The same procedure can be repeated when exploiting the generalized Callen-Suzuki spin identity~\cite{callen,Suzuki,balcerzak} for an exact calculation of the spontaneous magnetization of the  spin-3/2 magnetic ions C:
\begin{eqnarray}
m_{\mathrm{C}} \equiv  \langle \langle (1-n_j) S_k \rangle \rangle_c=(1-\rho)\left\langle\frac{\displaystyle\sum_{S_k=-3/2}^{3/2}S_k\mathrm{e}^{-\beta \mathcal{H}_k}}{\displaystyle\sum_{S_k=-3/2}^{3/2}\mathrm{e}^{-\beta \mathcal{H}_k}}\right\rangle=(1-\rho)\left\langle\frac{3\mathrm{e}^{\frac{9\beta D_{\mathrm{C}}}{4}}\sinh\left(\frac{3}{2}\beta J_{\mathrm{AC}}\zeta_k\right)+2\mathrm{e}^{\frac{\beta D_{\mathrm{C}}}{4}}\sinh\left(\frac{1}{2}\beta J_{\mathrm{AC}}\zeta_k\right)}   {2\mathrm{e}^{\frac{9\beta D_{\mathrm{C}}}{4}}\cosh\left(\frac{3}{2}\beta J_{\mathrm{AC}}\zeta_k\right)+2\mathrm{e}^{\frac{\beta D_{\mathrm{C}}}{4}}\cosh\left(\frac{1}{2}\beta J_{\mathrm{AC}}\zeta_k\right)}\right\rangle.
\end{eqnarray}
Following the same steps as described in above for the spontaneous magnetization of the ions B, one obtains the analogous expression for the spontaneous magnetization of the ions C:
\begin{eqnarray}
  m_{\mathrm{C}}= \frac{3}{2} (1-\rho) m_{\mathrm{A}}\left[g \left(\tfrac{3}{2}\right) +g\left(\tfrac{1}{2}\right)\right] +2 (1-\rho) t_{\mathrm{A}}\left[g\left(\tfrac{3}{2}\right)- 3g\left(\tfrac{1}{2}\right)\right],	
\end{eqnarray}
whereas the explicit form of the function $g(x)$ reads as follows:
\begin{eqnarray}
g(x) = \frac{3\mathrm{e}^{\frac{9\beta D_{\mathrm{C}}}{4}}\sinh\left(\frac{3}{2}\beta J_{\mathrm{AC}}x\right)+2\mathrm{e}^{\frac{\beta D_{\mathrm{C}}}{4}}\sinh\left(\frac{1}{2}\beta J_{\mathrm{AC}}x\right)}
  {2\mathrm{e}^{\frac{9\beta D_{\mathrm{C}}}{4}}\cosh\left(\frac{3}{2}\beta J_{\mathrm{AC}}x\right)+2\mathrm{e}^{\frac{\beta D_{\mathrm{C}}}{4}}\cosh\left(\frac{1}{2}\beta J_{\mathrm{AC}}x\right)}.
\end{eqnarray}
The spontaneous magnetization of the ions C can be alternatively expressed in terms of the parameters $V_3$ and $V_4$ given by Eqs. \eqref{v3}-\eqref{v4}:
\begin{align}
m_{\mathrm{C}}=\frac{3}{2} (1-\rho) m_{\mathrm{A}} \left(\frac{W_3}{V_3}+\frac{W_4}{V_4} \right)+2(1-\rho)t_A\left(\frac{W_3}{V_3}-\frac{3W_4}{V_4}\right)
\label{eq18}
\end{align}
and two newly defined functions:
\begin{align}
W_3=&3e^{\frac{9\beta D_{\mathrm{C}}}{4}}\sinh\left(\frac{9\beta J_{\mathrm{AC}}}{4}\right)+e^{\frac{\beta D_{\mathrm{C}}}{4}}\sinh\left(\frac{3\beta J_{\mathrm{AC}}}{4}\right), \label{eq21}\\
W_4=&3e^{\frac{9\beta D_{\mathrm{C}}}{4}}\sinh\left(\frac{3\beta J_{\mathrm{AC}}}{4}\right)+e^{\frac{\beta D_{\mathrm{C}}}{4}}\sinh\left(\frac{\beta J_{\mathrm{AC}}}{4}\right)\label{eq22}.
\end{align}
At this stage, one may easily calculate the total magnetization from the exact results \eqref{eq16}, \eqref{eq17} and \eqref{eq18} derived for the spontaneous magnetization of the ions A, B and C:
\begin{eqnarray}
m_t = m_{\mathrm{A}} + m_{\mathrm{B}} + m_{\mathrm{C}}.
\end{eqnarray}
The final formulas \eqref{eq17} and \eqref{eq18} for the spontaneous magnetization of the ions B and C are apparently odd functions of the coupling constants $J_{\mathrm{AB}}$ and $J_{\mathrm{AC}}$, which implies that a change of the ferromagnetic coupling constants ($J_{\mathrm{AB}}, J_{\mathrm{AC}} > 0$) to the antiferromagnetic ones ($J_{\mathrm{AB}}, J_{\mathrm{AC}} <0$) would merely cause a change of their relative orientation with respect to the spontaneous magnetization of the sublattice A. In the following, our particular attention will be restricted to the mixed-spin Ising model on a selectively disordered honeycomb lattice with a character of the ferro-ferrimagnetic ternary alloy AB$_{p}$C$_{1-p}$ with the antiferromagnetic nearest-neighbor coupling $J_{\mathrm{AB}}<0$ between the spin-1/2 and spin-1 magnetic ions, and respectively, the ferromagnetic nearest-neighbor coupling $J_{\mathrm{AC}}>0$ between the spin-1/2 and spin-3/2 magnetic ions. For simplicity, the relative size of the coupling constant $|J_{AB}|$ will be used for normalization purposes when defining two dimensionless quantities $J_{\mathrm{AC}} / |J_{\mathrm{AB}}|$ and $k_B T / |J_{\mathrm{AB}}|$ measuring a relative strength of the interaction ratio and temperature.

\begin{figure}
\includegraphics[scale=0.5]{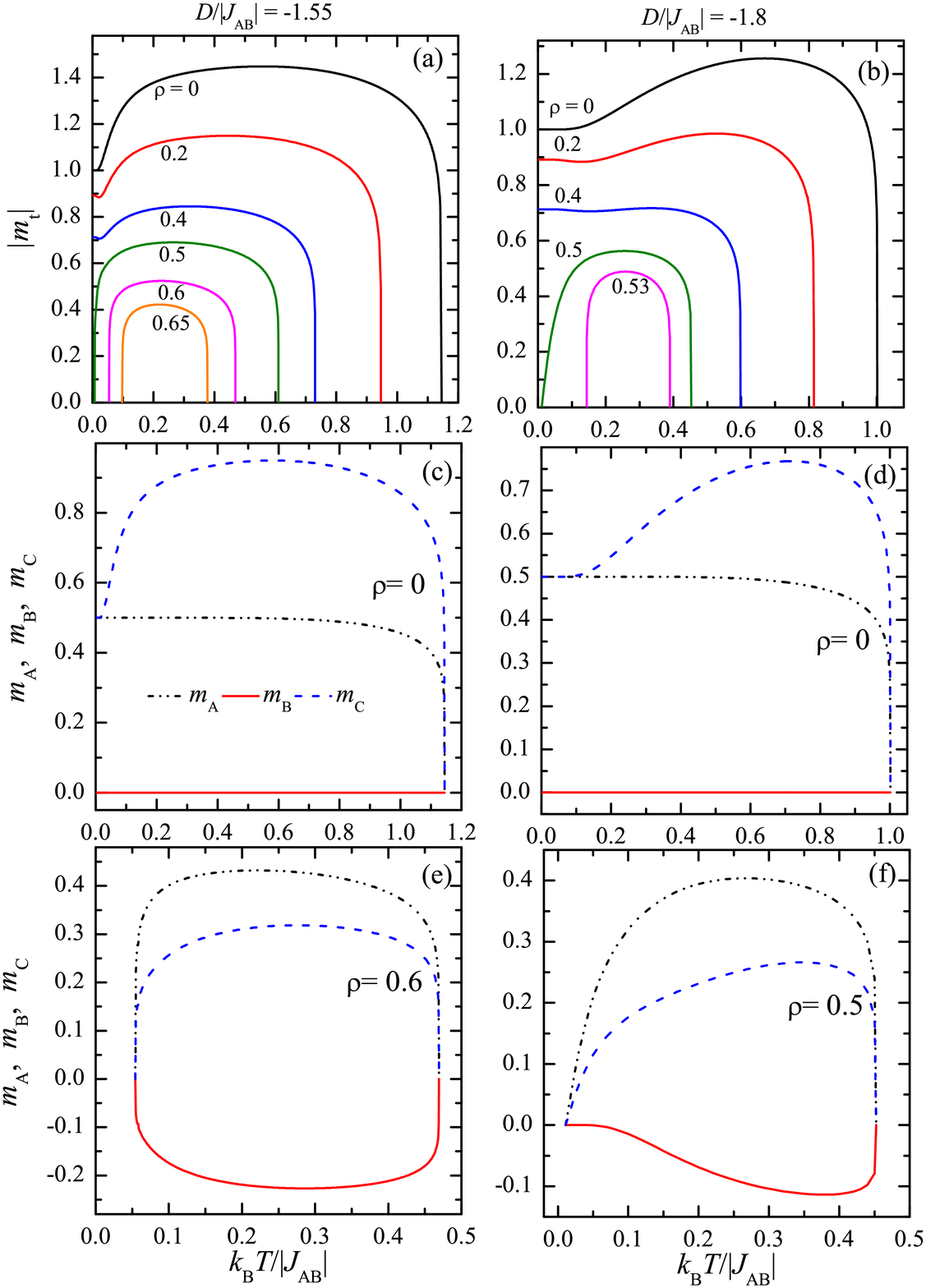}
\caption{\label{fig:5} Temperature variations of the total magnetization [Figs. \ref{fig:5}(a)-(b)] and the sublattice magnetization [Figs. \ref{fig:5}(c)-(f)] for the fixed value of the coupling ratio $J_{\mathrm{AC}}/|J_{\mathrm{AB}}| = 2$, several values of the concentration  of the spin-1 magnetic ions $\rho$ and two different values of the uniaxial single-ion anisotropy $D/|J_{\mathrm{AB}}| = -1.55$ (left panel)
and $-1.8$ (right panel).}
\end{figure}

First, let us comprehensively describe typical temperature variations of the spontaneous magnetization of the mixed-spin Ising model on a honeycomb lattice with the character of a ferro-ferrimagnetic ternary alloy for a specific value of the coupling ratio $J_{\mathrm{AC}}/|J_{\mathrm{AB}}| = 2$ assuming stronger ferromagnetic coupling $J_{\mathrm{AC}}$ in comparison with the antiferromagnetic one $|J_{\mathrm{AB}}|$. A few typical temperature dependencies of the total magnetization are plotted for this particular case in Figs. \ref{fig:5}(a)-(b) for several values of the concentration $\rho$ and two different values of the uniaxial single-ion anisotropy. For sufficiently low concentrations $\rho<0.5$, the total magnetization exhibits  a P-type temperature dependence when classified according to the standard N\'eel nomenclature \cite{neel}. Temperature variations of the sublattice magnetizations depicted in Figs. \ref{fig:5}(c)-(d) connect anomalous rise of the total magnetization upon increasing of temperature with a temperature-induced increase of the spontaneous magnetization of the ions C. The latter acts cooperatively with the spontaneous magnetization of the sublattice A due to the ferromagnetic character of the coupling constant $J_{\mathrm{AC}}>0$. Interestingly, for the concentrations $\rho \gtrsim 0.5$ and the uniaxial single-ion anisotropies $D/|J_{\mathrm{AB}}| \lesssim -1.5$, the total magnetization displays  a peculiar temperature dependence with double reentrant phase transitions [see Figs. \ref{fig:5}(a)-(b)], which goes beyond the standard N\'eel classification \cite{neel} and verifies the correctness of the established finite-temperature phase diagrams shown in Fig. \ref{fig:2}(a). To shed light on this unusual thermal dependence of the total magnetization, the sublattice magnetizations are plotted against temperature in Figs. \ref{fig:5}(e)-(f) for two special cases supporting the presence of the double reentrant phase transitions through the order-by-disorder effect \cite{vil80,she82}.

\begin{figure}
\includegraphics[scale=0.5]{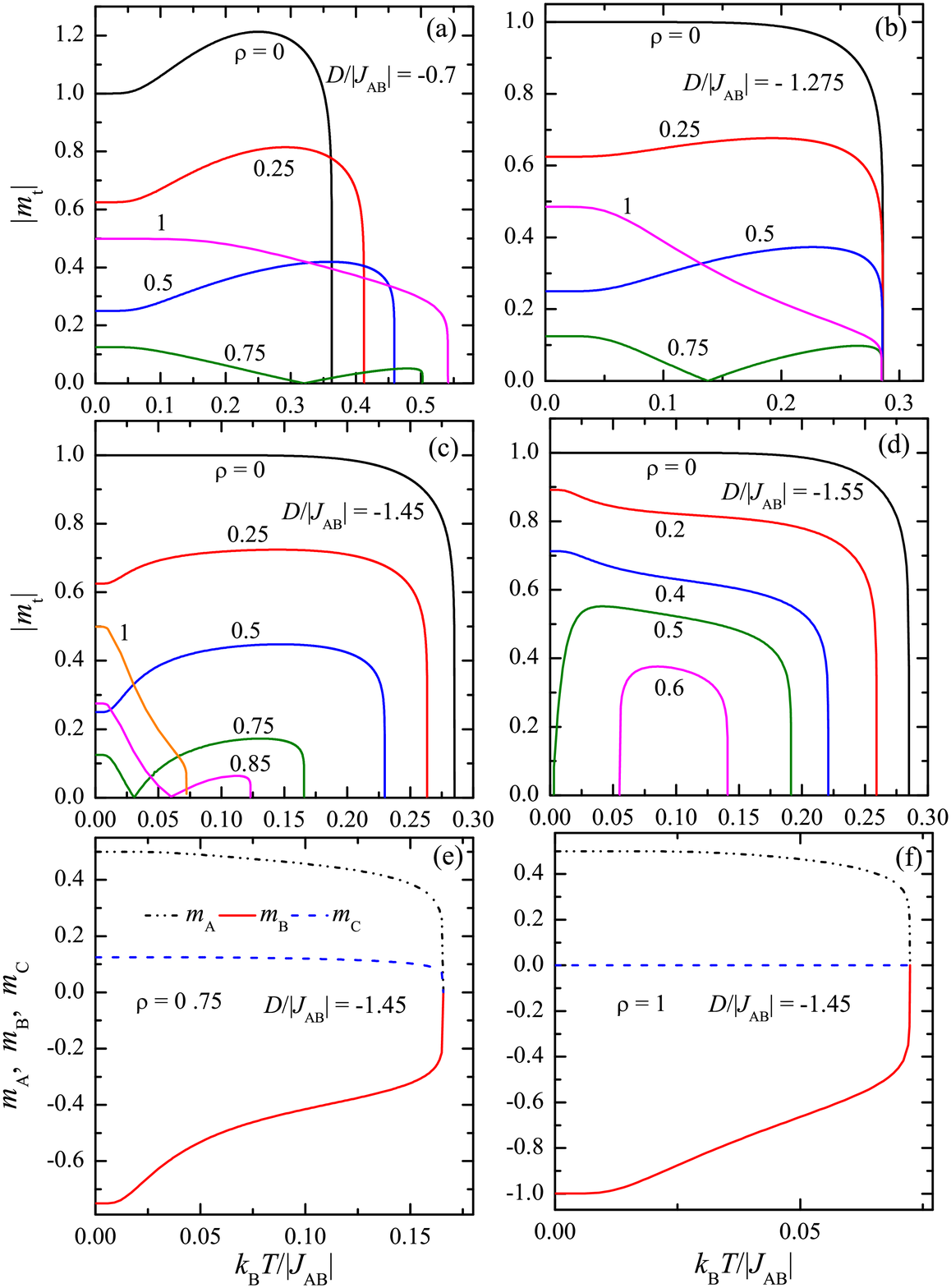}
\caption{\label{fig:6} Temperature variations of the total magnetization for the fixed value of the coupling ratio $J_{\mathrm{AC}}/|J_{\mathrm{AB}}| = 0.75$, several values of the concentration $\rho$ and four different values of the uniaxial single-ion anisotropy: (a) $D/|J_{\mathrm{AB}}| = -0.7$; (b) $D/|J_{\mathrm{AB}}| = -1.275$; (c) $D/|J_{\mathrm{AB}}| = -1.45$; (d) $D/|J_{\mathrm{AB}}| = -1.55$. Temperature variations of the sublattice magnetizations for the fixed value of the coupling ratio $J_{\mathrm{AC}}/|J_{\mathrm{AB}}| = 0.75$, the uniaxial single-ion anisotropy $D/|J_{\mathrm{AB}}| = -1.45$ and two different concentrations  of the spin-1 magnetic ions: (e) $\rho = 0.75$ and (f) $\rho = 1.0$.}
\end{figure}

Typical temperature dependencies of the spontaneous magnetization of the ferro-ferrimagnetic Ising ternary alloy on a selectively disordered honeycomb lattice are illustrated in Fig. \ref{fig:6}(a)-(d) for the particular case with the coupling ratio $J_{\mathrm{AC}}/|J_{\mathrm{AB}}| = 0.75$ assuming a weaker ferromagnetic coupling $J_{\mathrm{AC}}$ in comparison with the antiferromagnetic one $|J_{AB}|$. It directly follows from Fig. \ref{fig:6}(a)-(d) that the total magnetization displays a much greater diversity of temperature dependencies than in the reverse case. For instance, for the particular case with the uniaxial single-ion anisotropy $D/|J_{\mathrm{AB}}| = -1.45$, the total magnetization exhibits  the Q-type thermal dependence for the concentration $\rho = 0$, the P-type thermal dependence for the concentrations $\rho = 0.25$ and $0.5$, the N-type thermal dependence for the concentrations $\rho = 0.75$ and $0.85$, the R-type thermal dependence for the concentration $\rho = 1.0$  [see Fig. \ref{fig:6}(c)], all belonging to the standard classification scheme of ferrimagnets according to N\'eel theory \cite{neel}.  The most remarkable temperature variations of the total magnetization can be found for the concentration $\rho \approx 0.8$, since the total magnetization vanishes at a single compensation point below the critical temperature due to a complete cancellation of all three sublattice magnetizations in spite of the presence of a magnetic long-range order. The resulting N-type temperature dependence with a single compensation temperature follows from a change in sign of the total magnetization, which is negative below the compensation temperature due to the prevailing contribution of the sublattice magnetization $|m_{\mathrm{B}}| > m_{\mathrm{A}} + m_{\mathrm{C}}$ and positive above the compensation temperature because of the preponderant contribution of the sublattice magnetizations $m_{\mathrm{A}} + m_{\mathrm{C}} > |m_{\mathrm{B}}|$ [see Fig. \ref{fig:6}(e)]. It is also noteworthy that the R-type thermal dependence of the total magnetization observable in Fig. \ref{fig:6}(c) for the concentration $\rho = 1.0$ results from a relatively steep decline of the spontaneous magnetization of the sublattice B at low up to moderate temperatures, which is caused by a relatively strong value of the uniaxial single-ion anisotropy [see Fig. \ref{fig:6}(f)].

\begin{figure}
\includegraphics[scale=0.5]{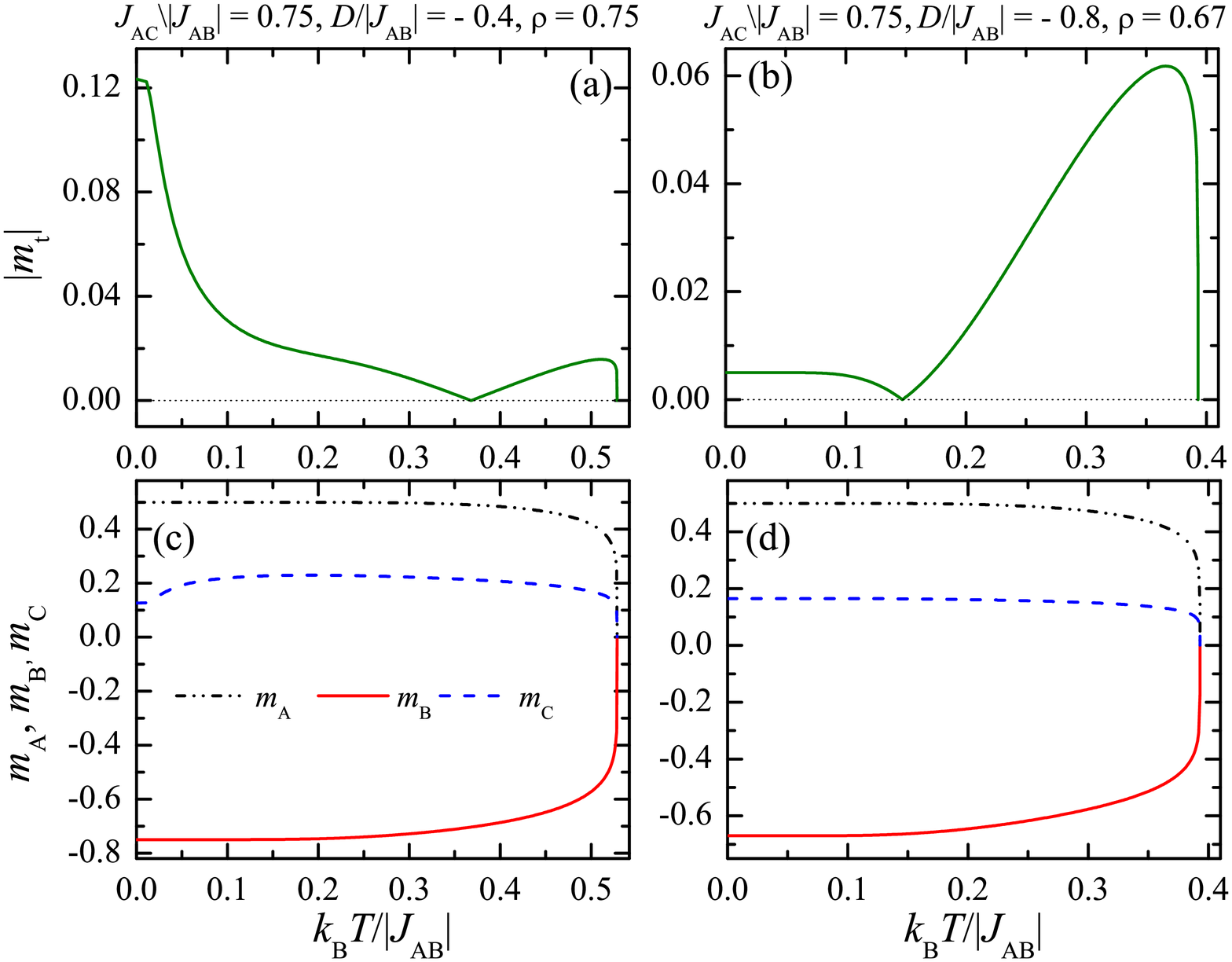}
\caption{\label{fig:7} Temperature dependencies of the total and sublattice magnetizations for a specific choice of the coupling ratio $J_{\mathrm{AC}}/|J_{\mathrm{AB}}|$, the uniaxial single-ion anisotropy $D/|J_{\mathrm{AB}}|$ and the concentration  of the spin-1 magnetic ions $\rho$, which lead to presence of a single compensation temperature.}
\end{figure}

Now, let us discuss in somewhat more detail N-type thermal dependencies of the spontaneous magnetization with a compensation temperature, which could be of technological relevance for a thermomagnetic recording. For this purpose, the total and sublattice magnetizations are plotted in Fig. \ref{fig:7}(a)-(d) against temperature for two different sets of the interaction parameters being compatible with a single compensation point. It is noteworthy that the relatively sudden drop of the total magnetization observable in Fig. \ref{fig:7}(a) at low enough temperatures can be connected to a thermally-induced rise of the sublattice magnetization $m_C$, which acts against the sublattice magnetization $m_B$ [Fig. \ref{fig:7}(c)]. On the other hand, the temperature variations of the total and sublattice magnetizations shown in Fig. \ref{fig:7}(b),(d) exemplify another outstanding N-type thermal dependence, which has greater absolute value of the total magnetization for temperatures exceeding a compensation point than below it.

\begin{figure}
\includegraphics[scale=0.5]{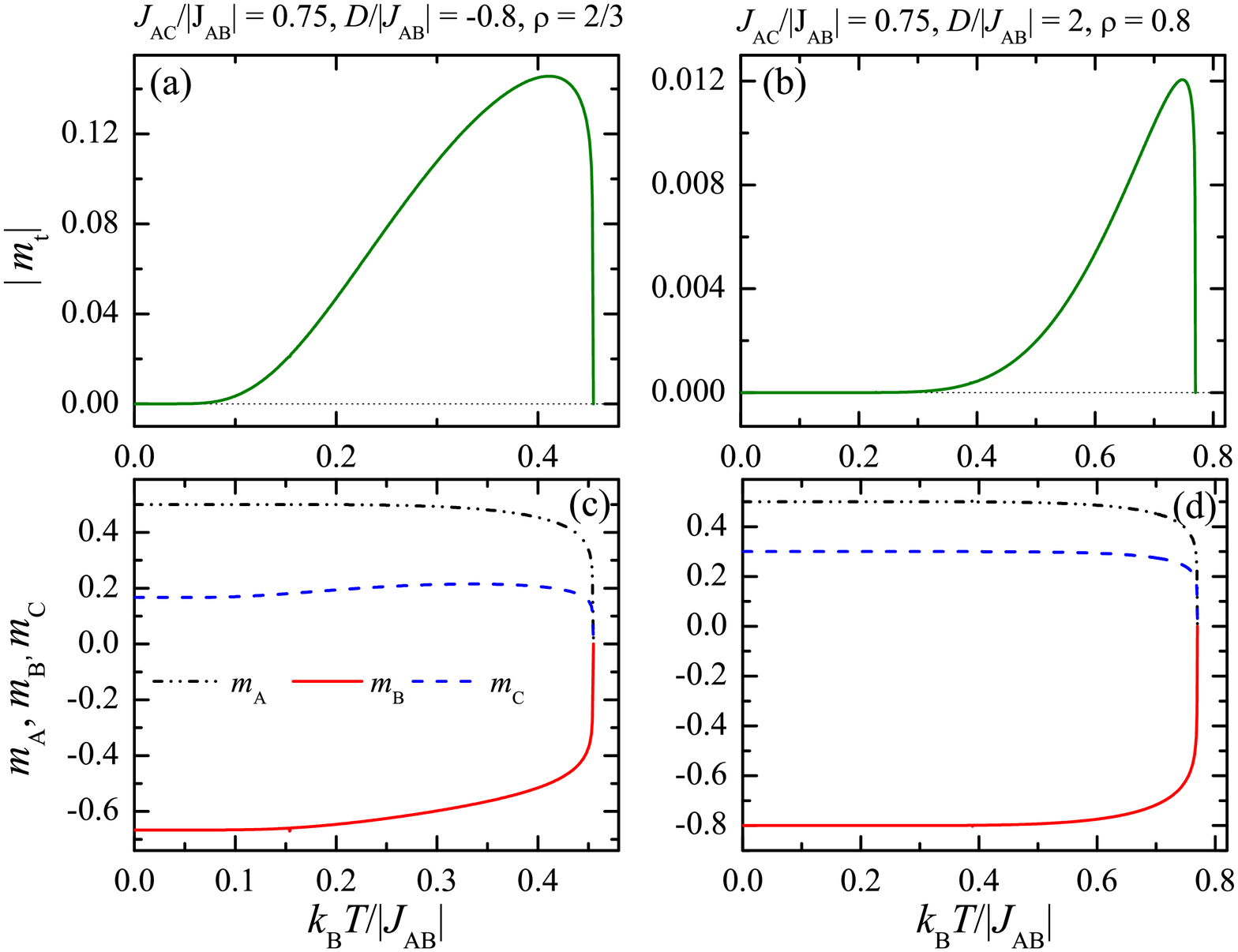}
\caption{\label{fig:8} Temperature dependencies of the total and sublattice magnetizations for a specific choice of the coupling ratio $J_{\mathrm{AC}}/|J_{\mathrm{AB}}|$, the uniaxial single-ion anisotropy $D/|J_{\mathrm{AB}}|$ and the concentration  of the spin-1 magnetic ions $\rho$, which lead to presence of a single compensation point in the asymptotic limit of zero temperature.}
\end{figure}

Last but not least, the total magnetization may also display the remarkable L-type temperature dependence, which can be regarded as a special case of the P-type dependence with an additional compensation point of the total magnetization emergent in the asymptotic limit of zero temperature \cite{str06}. This special compensation point can be simply achieved by a convenient choice of the concentration of the spin-1 and spin-3/2 magnetic ions. In general, there are two different possibilities how to achieve a full compensation of the three sublattice magnetizations at zero temperature. If the uniaxial single-ion anisotropy $D/|J_{\mathrm{AB}}|< -0.75$ ($D/|J_{\mathrm{AB}}|> -0.75$) forces the spin-3/2 magnetic ions to their lower (higher) spin state $S_j = 1/2$   ($S_j = 3/2$), then, the zero-temperature compensation points is reached for the specific value of the concentration $\rho = 2/3$ ($\rho = 4/5$), see Fig. \ref{fig:8}(a)-(d). Of course, the total magnetization starts to deviate from zero upon increasing of temperature due to different temperature variations of three individual sublattice magnetizations below the critical temperature shown in Fig. \ref{fig:8}(c)-(d).

\section{Conclusion}

In the present work we have investigated in detail phase diagrams and magnetization properties of a ferro-ferrimagnetic ternary alloy AB$_{\rho}$C$_{1-\rho}$, which is described in terms of the mixed spin-1/2, spin-1 and spin-3/2 Ising model on a honeycomb lattice with a selective site disorder on one of its sublattices. It has been demonstrated that the grand-canonical partition function of the investigated model can be rigorously mapped by means of a generalized star-triangle transformation to the canonical partition function of the effective spin-1/2 Ising model on a triangular lattice. Consequently, one may extract exact results also for the mixed-spin Ising model on a honeycomb lattice with a selective site disorder. It is also worthwhile to remark that the presented approach based on the generalized star-triangle mapping transformation can be rather straightforwardly adapted to the mixed-spin Ising models on other three-coordinated lattices as for instance a bathroom-tile or a square-hexagon-dodecagon lattice (see the sections 4.2.7 and 4.2.8 in Ref. \cite{aps}), but its application to the mixed-spin Ising models defined on lattices with different coordination number is not feasible.

It has been evidenced that the ferro-ferrimagnetic version of the studied model may display strikingly diverse temperature dependencies of the total magnetization, which may even include a compensation temperature for a specific choice of the relative concentration of the spin-1 and spin-3/2 magnetic ions. Besides, we provided a rigorous proof that the critical temperature of a ferro-ferrimagnetic ternary alloy may become independent of the concentration of the randomly mixed spin-1 and spin-3/2 magnetic ions for a specific value of the uniaxial single-ion anisotropy. The above features, namely, the appearance of compensation temperatures and of critical temperatures independent of the disorder concentration are in direct agreement with previous results gained from Monte Carlo simulations for a similar model of a ferro-ferrimgnetic ternary alloy on a square lattice with a quenched disorder restricted to a given sub-lattice \cite{gbuendia}. However, the most remarkable finding stemming from our present study is that the spontaneous ferro-ferrimagnetic ordering can be restored at finite temperatures through the order-by-disorder mechanism \cite{vil80,she82} in the parameter space $\rho \gtrsim 0.5$ and $D/|J_{AB}|\lesssim-1.5$ corresponding to a disordered (paramagnetic) ground state, which results in the anomalous temperature dependence of the total magnetization with double reentrant phase transitions.

\section*{Acknowledgments}
This work is dedicated to prof. Andrej Bob\'ak on the occasion of his 70th birthday anniversary. S. M. de Souza and O. Rojas acknowledge CNPq, CAPES and FAPEMIG. M. L. Lyra acknowledges CNPq, CAPES and FAPEAL. J. Torrico acknowledges CNPq (154080/2018-7). J. Stre\v{c}ka acknowledges the financial support by the grant of The Ministry of Education, Science, Research and Sport of the Slovak Republic under the contract No. VEGA 1/0531/19 and by the grant of the Slovak Research and Development Agency under the contracts Nos. APVV-14-0073 and APVV-18-0197.

\end{document}